\documentclass[10pt, conference, letterpaper]{IEEEtran}  
\usepackage{url,comment,cite}
\usepackage{amsmath,amssymb,amsfonts,bbm}
\usepackage{graphicx,color}
\usepackage{epstopdf}
\usepackage{epsfig,epsf,psfrag,textcomp,tabularx}
\usepackage[usenames,dvipsnames]{xcolor}
\usepackage{subfigure}
\usepackage{float}
\usepackage{stfloats,mathtools}
\usepackage[algoruled,medskip,dontprintsemicolon,linesnumbered,Algorithm]{algorithm}
\usepackage[noend]{algpseudocode}
\usepackage{float}
\usepackage{stfloats,mathtools}

\newcommand{\Fc}{\mathcal{F}}

\newcommand{\Lc}{\mathcal{L}}

\newcommand{\Nc}{\mathcal{N}}

\newcommand{\Sc}{\mathcal{S}}

\newcommand{\piv}{\boldsymbol{\pi}}

\numberwithin{equation}{section}

\begin{document}

\title{Energy-efficient Traffic Allocation in SDN-based Backhaul Networks: Theory and Implementation}

\author{\IEEEauthorblockN{Senay Semu Tadesse, Claudio Casetti, Carla Fabiana Chiasserini}
\IEEEauthorblockA{Politecnico di Torino, Italy} 
}

\maketitle
\thispagestyle{plain}
\pagestyle{plain}

\begin{abstract}
5G networks are expected to be highly energy efficient, with a 10
times lower consumption than today's systems. An effective way to
achieve such a goal is to act on the backhaul network by 
controlling the nodes operational state  and the allocation of traffic flows.  
To this end, in this paper we formulate 
energy-efficient flow routing on the backhaul network as an
optimization problem. In light of its complexity, which impairs
the solution in large-scale scenarios, we then propose a heuristic
approach. Our scheme, named EMMA, aims  to both turn off idle nodes and 
concentrate traffic on the smallest possible  set of
links, which in its turn increases the number of idle nodes. 
We implement EMMA  on top of ONOS and derive experimental results
by emulating the network through Mininet.
Our results show that EMMA provides excellent energy saving performance, which closely
approaches the optimum. In larger network scenarios,
the gain in energy consumption that EMMA provides with respect to the
simple benchmark where all nodes are active,  is extremely high, reaching almost 1 under 
medium-low  traffic load.  
\end{abstract}

\section{\label{sec:intro}Introduction} 

The 5G-Infrastructure-PPP (Public-Private Partnership), 
created to design and deliver architectures, technologies and standards 
for 5G communication, has set the following  KPIs (Key Performance Indicators) for 
energy management  \cite{5G-1,5G-2}: 
(i) energy efficiency improvement by at least a factor of 3 and
(ii) reduction of energy cost per bit by a factor of 10. 
Of course, no single solution can achieve such ambitious goals. Instead, they should 
be achieved through orchestrated actions, involving a fully-unified, automated
control and management plane, that oversees radio resources as well as computation
and transport resources in the the fronthaul and backhaul. A key role is played by
Software-Defined Networking (SDN) and Network Functions Virtualization (NFV),
tasked with the control and coordination of hundreds of nodes that need to be 
reconfigured on the fly in order to optimize utilization and QoS, in view of 
rapidly changing traffic flows. 
Among the coordinated actions that can be taken are the de-activation or 
decommissioning of scarcely used network portions, including links and switches,
and the flexible re-routing of existing flows so as to jointly address energy saving and 
QoE requirements.

In this paper, we address the latter action, by designing and evaluating an
Energy Monitoring and Management Application (EMMA), that can
minimize energy consumption of the backhaul network. 
Consistently with the pervasive use of SDN solutions expected in 5G
networks, EMMA natively interacts with the Northbound interface of ONOS~\cite{onos}, a popular
carrier-grade network operating system, whose 
Southbound interface is used to control OpenFlow switches. The design of EMMA hinges upon 
heuristic algorithms for the dynamic routing of flows and the 
management of the resulting link and switch activity. These algorithms  
represent a heuristic solution to a non-linear integer problem that aims
at minimizing 
the instantaneous power consumption of nodes and links. Performance 
evaluation has been done by comparing the optimum obtained through the above 
optimization formulation, with practical results derived 
by implementing the algorithms 
in an SDN network emulation environment.

The rest of the paper is organized as follows.
Section \ref{sec:problem} introduces the power
model for OpenFlow switches that we adopt and formalizes the problem
under study. Our heuristic scheme is presented in Section
\ref{sec:emma}, while its implementation on top of ONOS, as well as
the required interactions between ONOS and the underlying network, are described
in Section \ref{sec:implementation}. Emulation results and the
comparison between EMMA and the optimum solution are shown in Section
\ref{sec:peva}. 
A detailed discussion of previous work  and of our
novel contribution with respect to that, is provided in Section \ref{sec:rel-work}. 
Finally, Section \ref{sec:concl} draws some conclusions and
highlights future research directions.

\section{System Model and Problem Formulation\label{sec:problem}}
Energy consumption of the backhaul network can be minimized by limiting the number
of active links and nodes, i.e.,  by 
{\em (i)}  turning off link drivers whenever possible, resulting in
  proportional (possibly non-linear) changes, and 
{\em (ii)} turning off those nodes whose links are inactive. 

Both approaches can be studied by building a directed network graph whose
vertices represent the network nodes and edges correspond to  
links connecting the nodes. 
Let us then consider that the network includes $N$ nodes and $L$ links  and
denote by $\Nc$ and $\Lc$ the set of nodes and links,
respectively. Let a link $(i,j)\in \Lc$, with $i,j\in \Nc$, have a 
capacity $C(i,j)$ bits/s.  
Let $\Fc(t)$ denote the set of flows at time $t$, with each flow,
$f^{sd} \in \Fc(t)$, 
characterized by a source-destination 
pair,  a traffic volume, 
and  QoS constraints 
that in our case correspond to the required data rate $R(f^{sd})$. 

\begin{table}[th]
  \caption{Model notations\vspace{-3mm}}
  \label{tab:notation}
\begin{center}
  \begin{tabular}{||c|p{20mm}||c|p{20mm}|}
\hline \hline
$N$, $L$ & No. of nodes and links   &  $\Nc$, $\Lc$ & Set of nodes and links \\ \hline
$(i,j)\in \Lc$ & Link from node $i$ to node $j$ & $C(i,j)$ & Capacity of link $(i,j)$\\ \hline
$\Fc(t)$ & Set of active traffic flows at time $t$& $f^{sd} \in \Fc(t)$ & 
Active flow between source $s$ and destination $d$\\ \hline
$R(f^{sd})$ & Rate requirement for flow $f^{sd}$ &
$\piv$ & Generic path as an ordered sequence of links \\ \hline
$P_{idle}$ & Power consumption of an idle node    &
$P(i,j,t)$  & Power consumption associated with  link $(i,j)$ at $t$ \\\hline
   $x_{ij}(t)$ & Takes 1 if link $(i,j)$ is on at $t$, 0 else &
 $y_i(t)$ & Takes 1 if node $i$ is on at $t$, 0 else\\ \hline
$z_{\piv,f^{sd}}(t)$ &  Takes 1 if $f^{sd}$ is routed through 
path $\piv$ at $t$, 0 else & $\tau_{ij}(t)$ & Traffic  flowing over link $(i,j)$
at $t$ \\
\hline \hline
    \end{tabular}  
\end{center}
\end{table}

Let $x_{ij}(t)$ be a binary variable indicating whether link
$(i,j)\in \Lc$ is ``on'' ($x_{ij}(t)=1$) or ``off'' ($x_{ij}(t)=0$), at time $t$. 
Likewise,   $y_i(t)$ is a binary variable 
indicating whether node $i\in \Nc$ is active at time $t$ 
($y_i(t)=1$) or not ($y_i(t)=0$). Also, 
let a path $\piv$ be an ordered sequence of links. We indicate by the
binary variable $z_{\piv,f^{sd}}(t)$ 
whether flow $f^{sd}\in \Fc(t)$ is routed through 
path $\piv$ at time $t$ ($z_{\piv,f^{sd}}(t)=1$) or not ($z_{\piv,f^{sd}}(t)=0$). 

We consider that the generic node $i$ has zero
power consumption  when ``off'', and $P_{idle}$ when ``on'' but idle.  
The power consumption associated with a link,
$(i,j)$, at time $t$ linearly depends on the traffic that flows over
the link  and is denoted by 
$P(i,j,t)$. 
It follows that the total power
consumption of a node $i$ is given by:
\[P(i,t)= P_{idle}  +  \sum_{j \in \Nc, j \neq i} \left [P(i,j,t) x_{ij}(t)+ P(j,i,t)
  x_{ji}(t) \right ]\,.\] 
The traffic  flowing over 
link $(i,j)$
at time $t$, $\tau_{ij}(t)$, is expressed in bit/s and is 
given by the sum of the traffic associated with all flows which are
routed through the link, i.e.,
\begin{equation} \label{eq:tau}
\tau_{ij}(t) = \sum_{f^{sd} \in \Fc(t)} \sum_{\piv:(i,j) \in \piv}
R(f^{sd}) z_{\piv,f^{sd}}(t) \,.
\end{equation}

Below, we first present the power consumption model we adopt in order
to determine realistic values for  $P_{idle}$  and $P(i,j,t)$, for
OpenFlow switches. Then, we formalize the problem under study by using
 standard optimization.

\subsection{Power Model\label{subsec:power}}

The power consumption of an IP router or an Ethernet switch that is ``on'' is the sum of the power consumed by 
its three major subsystems \cite{Arun}:
 $P_{ctr} + P_{evn} + P_{data}$, 
where $P_{ctr}$ accounts for the power needed to manage the switch
and the routing functions, 
$P_{evn}$ is the power consumption of the 
environmental units (such as fans), and 
$P_{data}$ indicates the data plane power consumption. The latter can be decomposed into 
(i) a constant baseline component, and 
(ii) a traffic load dependent component. 
In other words, when a switch is powered on but it  does not  carry any data traffic,
 it consumes a constant baseline power.  
When a device is carrying traffic, it consumes 
additional load-dependent power for header processing, 
as well as for storing and forwarding the payload across the switch
fabric.
Combining the  power model in \cite{Arun} with that for OpenFlow
switches in \cite{Paul}, 
 we can write $P_{idle}$ as the sum of $P_{ctr}$, $P_{evn}$ and the base line component
of $P_{data}$, while the load-dependent component of $P_{data}$
is given by: 
\[P(i,j,t)=(E_{lookup}+ E_{rx}+E_{xfer}+E_{tx} )\tau_{ij}(t)\,.
\]
In the above expression,
\begin{itemize}
\item $E_{lookup}$ is the energy consumed per bit  in the
 {\em lookup} stage of a switch, which involves searching the TCAM for
 the received flow-key and retrieving the forwarding instructions;

\item $E_{rx}$ is the energy consumed per bit in the {\em reception} stage, which involves receiving a
  packet from the physical media, 
extracting important fields to build a flow-key and streaming 
 the packet into the input memory system;

\item $E_{xfer}$ is the energy consumed per
 bit in the {\em xfer} stage, which involves reading a
  packet from the inbound 
memory, all of the logic required to initiate a transfer across the 
fabric, driving the fabric connections and crossbar, as well as writing the packet into the remote outbound memory;

\item $E_{tx}$ is the energy consumed per
bit in the {\em transmission} stage, which involves reading a packet from 
the outbound memory and transmitting it on the physical media.
\end{itemize}
In the following, we set: $E_{rx}=E_{tx}=0.2$ nJ/bit, $E_{xfer}=0.21$ nJ/bit,
$E_{lookup}=0.034$ nJ/bit, and $P_{idle}=90$ W \cite{cisco-switches}.

\subsection{Minimum-energy Flow Routing\label{subsec:opt}}
The problem of energy-efficient flow allocation 
can be modeled similarly to what  done in \cite{Chiaraviglio,helle}.
However, we stress that, using the accurate power model introduced
above, optimal flow routing becomes a non-linear
integer problem (see our discussion at the end of this section). 
Furthermore, in order to achieve the minimum energy expenditure,
we aim at minimizing the {\em instantaneous} power consumption
of the network, i.e., the flow routing
problem should be solved whenever a new flow starts or an existing
flow ends. Typically, this is  impractical in
real-world networks but our goal here is to set the best performance
that could be achieved. 
We report below the formulation of the optimization problem adapted to our scenario.

{\bf Objective.} The goal is to minimize the instantaneous power
consumption, 
which is the sum of the power consumption due to the nodes being ``on''  and to active  link drivers:
\begin{equation}\label{eq:obj}
\min  \sum_{i \in \Nc} y_i(t) P_{idle}  + \sum_{(i,j) \in \Lc} x_{ij}(t) P(i,j,t) \,,
\end{equation}
where recall that $P(i,j,t)$ linearly depends on $\tau_{ij}(t)$, which
in its turn depends on the binary variable $z_{\piv,f^{sd}}(t)$, as
reported in Eq. (\ref{eq:tau}).

{\bf Constraints.}
 \begin{itemize}

\item Flow conservation constraint, for any $j\in \Nc$:
  \begin{eqnarray}\label{eq:flow-conservation}
\hspace{-3mm} && \hspace{-3mm} \sum_{i \in \Nc, i\ne j}\tau_{ij}(t) x_{ij}(t) - \sum_{k\in\Nc, k\ne
    j} \tau_{jk}(t) x_{jk}(t) \nonumber \\ 
\hspace{-3mm} & = & \hspace{-3mm} \sum_{f^{sd} \in \Fc(t): j=d}R(f^{sd})
  - \hspace{-3mm} \sum_{f^{sd} \in \Fc(t): j=s}R(f^{sd}) \,.
  \end{eqnarray}

 \item The total traffic flowing on a link must not exceed the link capacity:
 \begin{equation}\label{eq:limit-traffic}
  \tau_{ij}(t) \le C_{ij} \,,\,\,\,\forall (i,j) \in \Lc \,.
 \end{equation}

\item A link between two nodes, $i$ and $j$, can be active only if $i$
  and $j$ are both active: 
\begin{equation}\label{eq:ln-activation-constraint}
 \sum_{j\in \Nc, j\neq i} [x_{ij}(t) + x_{ji}(t)]\leq M y_i(t) \,, \,\,\,\forall
 i \in \Nc 
\end{equation}
where $M$ is an arbitrary  constant
s.t. $M \ge 2(N-1)$.


\end{itemize}

The input parameters of the above problem are 
the set of nodes, links and traffic flows, along
with their characteristics, while the decision variables are: $x_{ij}(t)$,
$y_{i}(t)$, and $z_{\piv,f^{sd}}(t)$. Thus, the problem is an integer
non-linear problem, due to the product of $x_{ij}(t)$'s and
$z_{\piv,f^{sd}}(t)$'s in the objective function and in the
flow-conservation 
constraints, which appears when 
$P(i,j,t)$ is expressed as a function of $\tau_{ij}(t)$, hence, of
$z_{\piv,f^{sd}}(t)$ (see Eq. (\ref{eq:tau})).
Also, it is akin to the bin packing 
problem\footnote{In the bin packing 
problem, objects of different volumes must be packed into a finite number of bins,
 each of a given capacity, in a way that the number
of used bins is minimized.}, which is a combinatorial 
NP-hard problem.  
Thus, obtaining the optimal problem
solution in large-scale scenarios is not viable.
Below we propose a heuristic algorithm that has low computational
complexity and whose performance results to be very close to the optimum.

\section{EMMA: A Heuristic Approach}\label{sec:emma}

In order to design an efficient algorithm to solve the above problem, we
first observe that an efficient heuristic for the solution of the bin
packing problem is  the First Fit algorithm \cite{vmplanner,gyarf}. 
Given the set of items to be inserted into the bins, the First Fit
algorithm processes an item at a time in arbitrary order and attempts to
place the item in the first bin that can accommodate it. If no bin is
found, it opens a new bin and puts the item in the new bin. 

We leverage the First Fit algorithm and design a heuristic
scheme, named Energy Monitoring and Management Application (EMMA), 
which: (i) monitors the network status, (ii) efficiently allocates traffic flows as
they come, and (iii) and re-routes
the existing flows when necessary and possible. Flows are (re-)routed
by EMMA with the aim to minimize the length of the flow path and the energy
consumption of the overall network. In particular, upon the arrival of
a new flow, EMMA first tries 
to fit the flow into the current ``active network'' while meeting the
flow traffic requirements. It then  turns on
other links and/or nodes only if  no suitable path
is found. 
Every time  a new link and/or node are added to the active network, 
EMMA looks for a better alternative path for 
flows that have been (re-)allocated long enough (more than half their
expected duration) ago. 
If a more energy-efficient allocation is found, then a flow is 
diverted on the new path, provided that its traffic  requirements
are still met. 
Note that EMMA differs from the First Fit  algorithm since it tries to find a better 
path for already allocated flows whenever any change in the active
topology occurs.

More in detail, the EMMA scheme is composed of two algorithms.
Algorithm~\ref{alg:newflow} presents the sequence of actions to be taken
whenever a new flow is activated in the network. Input to the algorithm
is the network topology and the information on the new flow to be
allocated, the power state of the network devices and the traffic
crossing every link. Initially, the computation of the possible
paths for the incoming flow is done considering the nodes and links that
are currently active (namely, the active network) (line 2). Then if
there exists a path that meets the flow traffic requirements, the flow
can be successfully allocated (lines 3-5). Otherwise, the whole network
should be considered and the search for a suitable path repeated (lines
6-7). If a path is found, the links and nodes that need to be added
are activated 
(lines 8-11). 

\begin{algorithm} \caption{New flow allocation\label{alg:newflow}}
{\small 
\begin{algorithmic}[1]
\Require Topology, new flow, network power state, traffic load
\For {each new flow}
\State {Compute all shortest paths across active topology}
\If {suitable path is found}
\State \% if more than one, select one at random
    \State Allocate the flow
\Else 
    \State {Compute shortest paths considering the whole network}
    \If {suitable path is found} 
\State \% if more than one, select one at random
        \State Turn on the selected links and nodes that are off
        \State Allocate the flow 
        \State installation\_time $\leftarrow$ current\_time
        \State Run Algorithm~\ref{alg:reroute} 
\State \% It moves previous flows to a
        better path if any
    \EndIf    
\EndIf
\EndFor
\end{algorithmic}
}
\end{algorithm}

Algorithm~\ref{alg:reroute} states the steps followed during re-routing of existing traffic. 
This algorithm is run whenever there is a change in 
the active network topology, i.e., if nodes and/or links become active while finding a path for a new flow.

Input to the algorithm are the network topology and the information on
the current flows. 
The algorithm  selects all flows that have started or re-allocated 
at least $T_a$ time ago (line 1). For each flow satisfying the time
hysteresis, 
it computes a path on the current topology starting with the flows 
with higher rate requirements  (lines 2-4). If the cost
of the computed path is less than that of the flow current path, it diverts 
the flow to the new path and updates the flow
installation time (lines 5-7). If the process of moving flows to a
different path results in some links and/ or nodes being idle, 
those links and/or nodes are turned off (lines 8-9).

\begin{algorithm} 
\caption{Move flows to a better path\label{alg:reroute}} 
{\small 
\begin{algorithmic}[1]
\Require Topology, information on current flows 
\State $\Sc \leftarrow \{\mbox{\rm flows s.t. installation\_time $\geq$ $T_a$}\}$ 
\State Order flows in $\Sc$ with decreasing rate requirements
\For {each flow in $\Sc$}
    \State {Search a path on the active topology}
\If {suitable path exists $\wedge$  (new path cost $<$ old path cost)}
    \State Move  flow from old to new path 
    \State installation\_time $\leftarrow$ current\_time
\EndIf
\EndFor
\If {there are links and/or nodes no longer carrying traffic}
    \State {Turn them off}
\EndIf
\end{algorithmic}
}
\end{algorithm}

\begin{figure}
\centering
\includegraphics[width=0.48\textwidth]{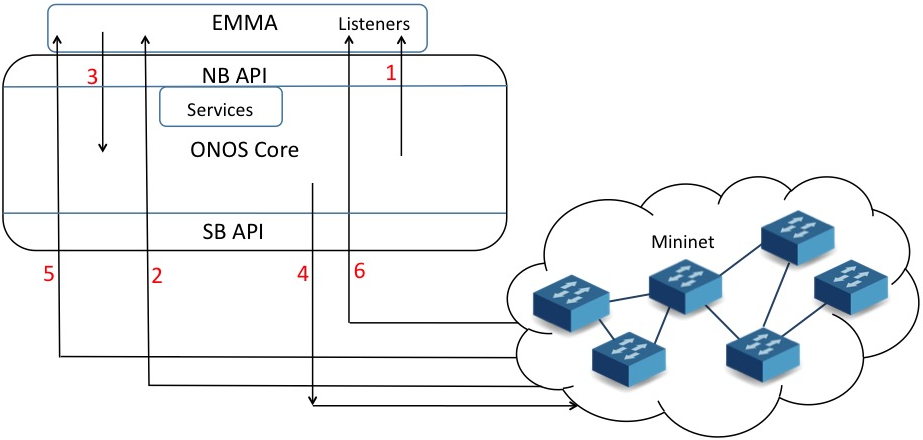}
\caption{\label{fig:implementation}Interaction between EMMA and ONOS,
  and between ONOS and Mininet. The interactions are labeled with
  numbers corresponding to the sequence of events described in the text.}
\end{figure}

\section{EMMA Implementation in ONOS\label{sec:implementation}}

We consider a system architecture including:
\begin{itemize} 
\item a network of OpenFlow switches and links interconnecting them;
\item an SDN controller;
\item an application on top of the SDN controller that implements EMMA.
\end{itemize}
The system has to:
\begin{itemize}
\item
keep track of the set of active network elements, and the set of current traffic flow routes;
\item
route traffic (i.e., check routing paths for traffic flows and push routes into the network);
\item
monitor and toggle the power state of the network elements.
\end{itemize}

In SDN,  traffic forwarding rules are created by
the controller and are installed into the switches. The flow rule
installation could be either proactive, e.g., flows rules are installed into the
switches prior to the actual flow arrival, or reactive, e.g., 
flows rules are created upon every new flow coming into the switches. 
In the latter case, the switches at the network edge will need to be
always active, in order to receive the traffic that a host may
generate. When an edge switch receives a packet for which it has no
rule, it directly sends the packet to the
controller. Then the
controller gets from  EMMA  a path for such kind of packets and installs a
flow rule into the switches. Afterwards, all packets of that kind will be
forwarded based on the flow rule just installed.

In the implementation of our algorithms we followed a reactive
approach. The algorithms are tested using the Mininet SDN network
emulation environment \cite{Mininet} as forwarding plane, and a controller whose
functions are supported by the Open Network Operating
System (ONOS) \cite{onos}. The algorithm is developed based on
ONOS built-in Reactive Forwarding application, which uses
the shortest path to route incoming traffic. 
However, we recall that the topology considered for the shortest
path search  is not always the whole topology. Indeed, EMMA  first
tries to find a path on the nodes/links which are already carrying
traffic 
thus concentrating all the traffic in the active infrastructure whenever possible. 
If no path is found using the active network, EMMA
considers the whole network to route the traffic.

Also, it is worth mentioning here 
that EMMA implements the {\em PacketProcessor} interface of ONOS,
running below it in the architecture,
in order to process packets, i.e., we replicated the processing
function at the {\em PacketProcessor} interface of ONOS within EMMA.
In this way, EMMA can distinguish between traffic and control packets,
as well as between unicast and multicast packets, thus further
processing only unicast traffic packets. 

Finally, we consider that initially all network nodes and links are ``on''
   and that ONOS provides  EMMA with the whole network
 topology. Specifically, EMMA gets the network topology via the {\em TopologyService}
 application of ONOS, to which EMMA has registered.
EMMA can then use the network topology for path computation and flow provisioning.
If there are idle core switches, EMMA will instruct ONOS to turn them
off.



Figure \ref{fig:implementation} highlights the main components, 
 chain of events and interactions in the implemented system. 
The description of the set of actions taken by  
EMMA and the list of events happening in the network are detailed 
below (the labels associated with the arrows in the figure correspond 
to the numbers of the listed events/actions).


\begin{enumerate}
 \item 
Whenever a switch receives a packet for which it does not have a
   forwarding rule, 
it sends the packet of the flow directly to ONOS. 
EMMA intercepts the packet using the {\em requestPacket} method of
the {\em PacketService} interface of ONOS.  

 \item  EMMA processes the packet and, if it carries unicast traffic,
it computes a flow path according to Algorithm \ref{alg:newflow}. 
Then it creates a forwarding rule using a set of 
ONOS applications. Initially, it 
uses {\em TrafficSelector.builder}, to 
form the part of the rule specifying the pair of source and
destination IP addresses to be matched when processing the packet at
the switch. 
It then resorts to {\em TrafficTreatment.builder} in order to express a
set of instructions, namely, forwarding the packet toward the
intended output port and decrementing the packet TTL. 
Finally, it invokes {\em ForwardingObjective.Builder} to
create the forwarding rule based on the {\em TrafficSelector} and
{\em TrafficTreatment} outputs.

 \item The ONOS application {\em FlowObjectiveService} is used to send
the rule and install it in the switch.

 \item A FLOW\_ADDED event will be generated by the  
 switch after the flow rule has been installed. Following 
the blueprint of the {\em FlowListener}
interface of ONOS, we built a {\em FlowListener}
interface within EMMA. Such an interface allows EMMA to detect a
FLOW\_ADDED event, after which EMMA starts accounting for 
the power consumption due to the newly added flow. 

\item A FLOW\_REMOVED event will be generated by a switch 
 whenever a flow finishes, or the application removes a flow from a
 switch when a better path is found. As for the  FLOW\_ADDED event,
 EMMA is able to detect a flow removal through the {\em FlowListener}
interface we developed and, hence, to correctly compute the energy
consumption of the active network. Note that when a FLOW\_REMOVED event
corresponds to a flow termination (i.e., it has not been triggered by
an EMMA re-routing action), EMMA will check whether existing flows
can be re-routed, according to Algorithm \ref{alg:reroute}, thus
performing a further step similar to step 2). 
\end{enumerate}

We conclude this section by remarking that in our implementation 
 EMMA computes the active network power
consumption based on the packet size and the flow rate, as explained
in Section \ref{subsec:power}. 
Such values are obtained as 
the ratio of, respectively,  the number of bytes to the number of packets, and
 the number of packets to the flow duration. The number of packets,
number of bytes and flow duration are provided to EMMA
by ONOS. 
Specifically, EMMA exploits the {\em getDevice} method of {\em DeviceService} in
ONOS to get the list of available   switches, and the  {\em getFlowEntries}
method of the {\em FlowRuleService} of ONOS to get information (i.e., 
number of bytes, number of packets and flow duration) about each flow
handled by a given switch.


\begin{table}
  \caption{Default settings\vspace{-3mm}}
  \label{tab:3}
\begin{center}
  \begin{tabular}{ | l | l |}
    \hline
    Parameter & Value \\ \hline \hline
    Flow arrival rate  & 0.1 flows/s \\ \hline
    Average flow duration & 20 s\\ \hline
    Number of core switches & 12 \\ \hline
    Number of edge switches  & half the no. of core switches \\ \hline
    Link Capacity & 10 MB/s \\ \hline
    Hysteresis & 10 s \\ \hline
    $P_{idle}$ & 90 W \cite{cisco-switches}\\ \hline
    $P(i,j,t)$ &  $0.644 \cdot \tau_{i,j}(t)$ nW \cite{cisco-switches}\\ \hline
    Number of hosts per edge switch & 10 \\ \hline
    Link prob. b/w switches & 0.5 \\ \hline  
    Packet size & 1500 bytes \\ \hline 
    Experiment duration & 500 s\\ \hline
  \end{tabular}
\end{center}
\end{table}

\section{Experimental Results\label{sec:peva}}

We evaluated the EMMA performance against the optimal solution, as
well as versus the simple case where no power saving strategy is
adopted and the whole network is always active (hereinafter referred
to as No Power Saving). 
The performance of EMMA and of the No Power Saving schemes are
obtained by emulation, in the system we implemented and that is
described above. 
The solution of the optimization problem in Eq. (\ref{eq:obj}) is instead obtained
using the Gurobi solver, considering the same network as that emulated 
in our experiments with Mininet. 

We derived the results assuming a default number of core and edge
switches equal to 12 and 6, respectively;
10 hosts are connected to each edge switch. Links between any two
core switches are set with probability 0.5 and the
link capacity is set to 10 Mbytes/s. TCP traffic flows are generated
using the Iperf tool, using 1500-byte packets.
For each traffic flow, source and  destination are selected at random among all
possible hosts. Note that this is a worst case assumption for EMMA,
while it favours the No Power Saving strategy.
The inter-arrival time of newly generated flows follows a negative
exponential distribution with a default mean arrival rate of 0.1
flows/s. The traffic flow duration is also  exponentially distributed
with mean equal to 20 s.
The complete list of default values that we adopted for the system
parameters is reported in Table \ref{tab:3}.

In the following figures, we show the average power consumption per
flow, as the flow arrival rate and the number of network switches vary.
The results have been obtained by averaging over 20 experiments. 
Note also that power consumption is computed based on traffic
statistics and nodes operational states, consistently in all cases.

\begin{figure}
\centering
\includegraphics[width=0.45\textwidth]{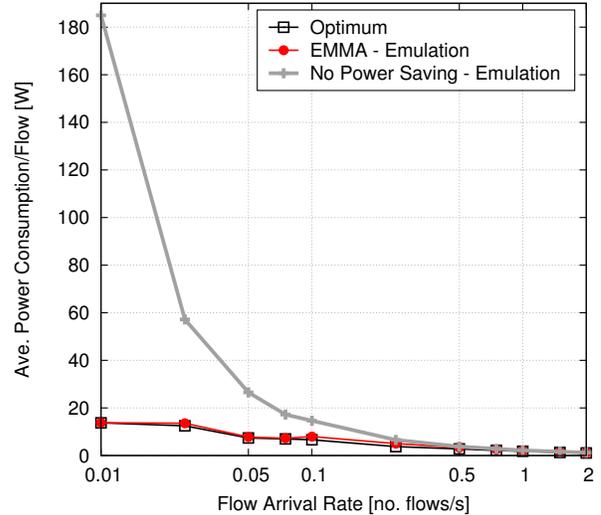}
\caption{Comparing EMMA against the optimum and the No Power Saving
  scheme: Average power consumption per flow as a function of the flow
  arrival rate (no. core switches = 12).\label{fig:vslambda}}
\vspace{-5mm}
\end{figure}

\begin{figure}
\centering
\includegraphics[width=0.45\textwidth]{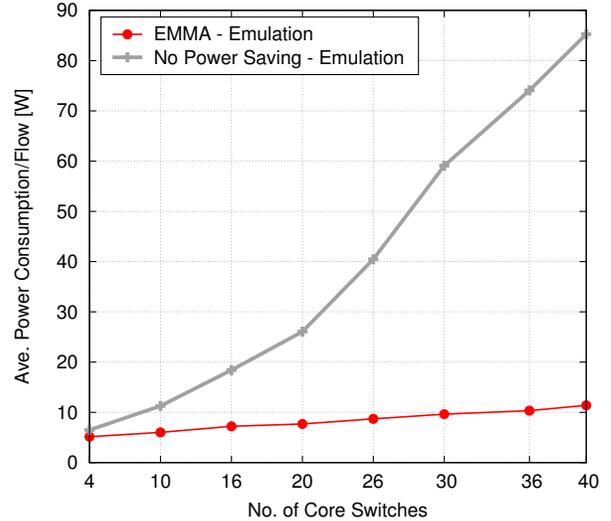}
\caption{Average power consumption per flow vs.  number of core
  switches: comparison between EMMA and No Power Saving (flow
  arrival rate = 0.1)\label{fig:vsN}}
\vspace{-5mm}
\end{figure}

\begin{figure}
\centering
\includegraphics[width=0.45\textwidth]{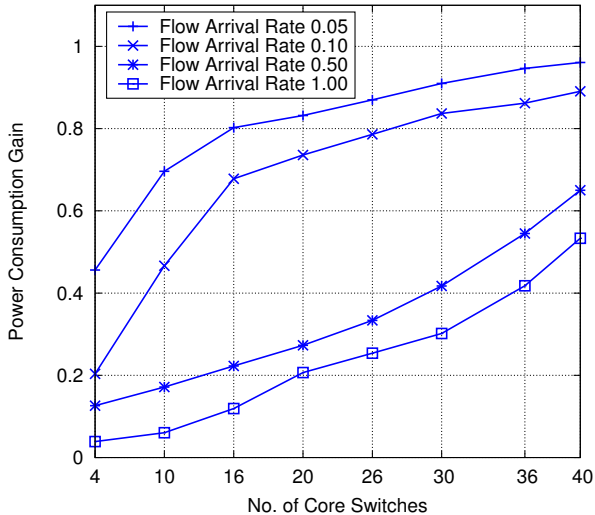}
\caption{Gain in average power consumption per flow (derived by
  emulation) provided by EMMA
  with respect to No Power Saving, as the number of core
  swicthes and the flow arrival rate vary.\label{fig:all}}
\vspace{-5mm}
\end{figure}

Figure \ref{fig:vslambda} compares the performance of EMMA to the
optimum as well as to that of the No Power Saving scheme, as the flow
arrival rate varies and for the default number of core switches
(namely, 12). Observe that EMMA matches the optimum very closely,
for any value of flow arrival rate. The power saving it provides with respect to
the case where all network switches are always on is very
noticeable. Clearly, the power gain tends to shrink when many flows
have to be allocated 
(high flow arrival rate), i.e., as 
an increasing number of switches and links have to be used. 

The behavior of EMMA compared to the No Power Saving scheme, as the
network size varies, is presented in
Figure \ref{fig:vsN}. Here we do not show the optimum performance,
as we could not solve the optimization problem for a number of core
switches significantly larger than the default value.
All results are therefore derived by emulation,
under a flow arrival rate equal to 0.1. The plot confirms the
excellent performance of EMMA: it reduces the power
consumption per flow by a factor ranging from 2 (for 10 core switches)
to 8 (for 40 core switches). As noted before, a smaller improvement is obtained only when 
the network size is small compared to the flow arrival rate  (e.g.,
for 4-5 core switches).

Finally, Figure \ref{fig:all} depicts 
the gain that we can achieve with EMMA with respect to 
the No Power Saving strategy, as a function of the number of core switches
and for a flow arrival rate equal to 0.05, 0.1, 0.5, 1. 
The gain is computed as the difference in power consumption between
No Power Saving  and EMMA, normalized to the power consumption of the
former scheme. 
As expected, 
the gain that EMMA provides is higher for a lower value of flow
arrival rate 
and a larger network size, since it is possible to aggregate more
flows on the same links and there are more idle switches that
can be turned off. Interestingly, the gain we obtain is always quite
high, with peak values that approximate 1.

\section{\label{sec:rel-work}Related Work} 
Energy-efficient 
traffic routing in wired networks has been largely addressed in the
literature. Here, we limit our discussion to the
studies that are most relevant to
ours. In particular, at the end of the section we highlight our major contributions
with respect to those that are closest to our study.  

One of the first works to investigate energy-efficient management of
nodes and links in backbone networks, can be found in \cite{Chiaraviglio}. The
optimization problem they present to obtain an energy-efficient
traffic allocation is an integer linear problem ILP, thus  a heuristic is
proposed too. Their algorithm first turns off nodes
with the smallest traffic load and re-routes traffic consequently,
then it tries to de-activate links. 
An opposite approach with respect to \cite{Chiaraviglio} is adopted in
\cite{radu,frederic}, where the least congested links are turned off first.

In \cite{vmplanner},  both virtual
machine (VM) placement and traffic flow routing are optimized so as to turn
off as many unneeded network elements as possible.  
In particular, the authors use traffic-aware VM grouping to partition
VMs into a set of VM-groups so that the
overall inter-group traffic volume is minimized while the overall 
intra-group traffic volume is maximized. 
An approach based on the greedy bin-packing algorithm is proposed 
to route the traffic, and to put as many network elements as possible into sleep mode.
A similar approach is adopted in \cite{okonor}, which focuses on the
case where a sudden surge in traffic occurs after an off-peak period,
during which most of the nodes have been turned off. 
The work tries to minimize the number of nodes/links that have to be
activated and to reduce service disruption by avoiding turning off
links that are critical to guaranteeing network connectivity. 

Relevant to our work are also the studies in 
\cite{hero,helle}, which  minimize the power consumption of a data center
network. 
In particular, 
\cite{hero} considers a hierarchical topology and proposes a
hierarchical energy optimization technique: 
all edge switches connecting to any source or destination server in the
traffic matrix must be ``on''. Also, the network is
divided into several pod-level subnetworks and a core-level
subnetwork, and traffic is re-organized accordingly. 
\cite{helle} instead assumes that each traffic flow can
be split over different paths. Interestingly, they have used OpenFlow
to collect the flow matrix and port 
counters, which are used as input to their routing scheme.  
The work in \cite{tran} extends \cite{helle} by introducing a monitoring module that collects
statistics, such as switch state, link state, active topology and traffic utilization. 

Finally, physical characteristics of the links are
accounted for in \cite{fisher,angelo}. The former considers
network routers connected by multiple physical cables forming one logical
bundled link, and it aims at turning off the cables within such links. 
The problem  \cite{fisher} poses accounts for the bundle
size, besides the network topology and the traffic matrix, and it
consists in maximizing the spare network capacity by minimizing the
sum of loads over all links. Instead, in \cite{angelo} the focus in on
optical links and the 
minimum-energy traffic allocation is solved taking into account their peculiarities.

We remark that \cite{Chiaraviglio,vmplanner} are the closest works to
ours, nevertheless the study we present significantly differs from them. In
particular, we recall that  our problem
formulation resembles that in \cite{Chiaraviglio}, but it accounts
for the instantaneous power consumption and for a more realistic model 
of the nodes power consumption, which 
changes the nature of the optimzation. As far as  our heuristic is
concerned, 
we leverage \cite{vmplanner} but design an algorithm that, unlike \cite{vmplanner},  aims to
 find a better route for all existing flows whenever there is a change
 in the active topology. In addition, our focus is on the
 implementation of energy-efficient flow routing: 
our algorithms are implemented on ONOS, we define and implement the
interactions that our application requires between ONOS and the network emulated through
Mininet,  and we derive emulation results by letting ONOS and Mininet interact.

\section{Conclusions\label{sec:concl}} 
We addressed energy-efficient flow allocation in the 5G backhaul 
where traffic forwarding rules are created by
an SDN controller, which can also turn switches on or off.
We first formalized flow allocation by formulating an
optimization problem whose complexity, however, results to be
unbearable in large-scale scenarios. We therefore used a
heuristic approach and developed an application, named EMMA.
We implemented EMMA on top of ONOS and derived experimental results by
emulating the network through Mininet. The comparison between EMMA and
the optimal solution (obtained in a small-scale scenario) showed that the EMMA  
performance is very close to the optimum. Also, in larger scale scenarios, emulation results
highlighted that EMMA can provide a dramatic energy improvement 
with respect to our benchmark where switches are always on.

Future research will address the problem of energy-efficient
allocation and migration of virtual machines, and it will combine the
solution with the EMMA scheme proposed in this paper for flow routing.

\section*{Acknowledgment}
This work has received funding from the 5G-Crosshaul project (H2020-671598).
The authors would like to thank Dr. Giada Landi (Nextworks) for the helpful
 discussions.


\end{document}